\documentclass[pdflatex,sn-apa]{sn-jnl}

\usepackage{graphicx}%
\usepackage{multirow}%
\usepackage{amsmath,amssymb,amsfonts}%
\usepackage{amsthm}%
\usepackage{mathrsfs}%
\usepackage[title]{appendix}%
\usepackage{xcolor}%
\usepackage{textcomp}%
\usepackage{manyfoot}%
\usepackage{booktabs}%
\usepackage{algorithm}%
\usepackage{algorithmicx}%
\usepackage{algpseudocode}%
\usepackage{listings}%
\usepackage{enumitem}%

\theoremstyle{thmstyleone}%
\theoremstyle{thmstyletwo}%

\theoremstyle{thmstylethree}%

\begin{document}

\title[Article Title]{SHAPR: Operationalising Human–AI Collaborative Research Through Structured Knowledge Generation}


\author{\fnm{Ka Ching} \sur{Chan}}\email{kc.chan@unisq.edu.au}




\affil*[1]{\orgdiv{School of Business}, \orgname{University of Southern Queensland}, \city{Springfield Central}, \postcode{QLD 4300}, \state{Queensland}, \country{Australia}}




\abstract{SHAPR (Solo Human-Centred and AI-Assisted Practice) is a framework for research software development that integrates human-centred decision-making with AI-assisted capabilities. While prior work introduced SHAPR as a conceptual framework, this paper focuses on its operationalisation as a structured, traceable, and knowledge-generating approach to AI-assisted research practice. We present a set of interconnected models describing how research activities are organised through iterative cycles (Explore–Build–Use–Evaluate–Learn), how artefacts evolve through development and use, and how empirical evidence is systematically transformed into conceptual knowledge. Central to this process is the notion of Structured Knowledge Units (SKUs), which provide modular and reusable representations of insights derived from practice, supporting knowledge accumulation and reuse across cycles. The framework further introduces evidence and traceability as a cross-cutting mechanism linking human decisions, AI-assisted development, artefact evolution, and resulting knowledge, enabling transparency, reproducibility, and systematic refinement. SHAPR is also positioned as an AI-executable research framework, as its structured processes and documentation practices can be interpreted and enacted by generative AI systems to guide research workflows. At the same time, SHAPR supports a continuum of AI involvement, allowing researchers to balance control, learning, and automation across different research contexts. Beyond individual workflows, SHAPR is conceptualised as an integrated research system that combines LLM workspaces, development environments, cloud storage, and version control to support scalable, traceable, and knowledge-centred research practices. Overall, SHAPR provides a practical and theoretically grounded foundation for conducting rigorous, transparent, and reproducible research in AI-assisted environments, contributing to the development of scalable and methodologically sound research practices.}




\maketitle

\section{Introduction}\label{2intro}

The rapid emergence of generative artificial intelligence has significantly transformed how researchers design, prototype, and evaluate digital artefacts. Large language models and AI-assisted development tools are increasingly capable of generating code, proposing architectural designs, and supporting debugging and experimentation. These developments reflect a broader shift toward AI-augmented research environments, where human and machine capabilities are increasingly integrated \cite{Shneiderman2022}. These capabilities enable researchers to develop complex research software artefacts more rapidly than ever before. However, they also introduce new methodological challenges related to the structure, traceability, and reproducibility of AI-assisted research workflows.

In many contemporary research environments, substantial portions of reasoning, experimentation, and design exploration occur within conversational interactions between researchers and generative AI systems. While these interaction environments can accelerate development, they also make it difficult to document how design decisions are made and how research knowledge emerges from iterative experimentation. Without a structured approach, AI-assisted development risks becoming opaque, difficult to replicate, and challenging to evaluate within a rigorous research process.

Design Science Research (DSR) and Action Design Research (ADR) provide well-established foundations for artefact-based knowledge generation, emphasising iterative development and evaluation to produce design knowledge \cite{Hevner2004, Peffers2007, Sein2011}. However, these methodologies were developed in contexts where software development was typically conducted by teams using conventional workflows. They provide limited operational guidance for researchers who increasingly conduct artefact development individually while interacting extensively with generative AI systems.

This creates a methodological gap: researchers require structured approaches for managing AI-assisted, often solo, research software development in a manner that remains traceable, reproducible, and methodologically rigorous. Addressing this gap is particularly important as AI tools become more capable of generating complex artefacts. While generative AI can augment development activities, research remains fundamentally a human epistemic process in which the researcher retains responsibility for interpretation, evaluation, and knowledge claims.

The SHAPR framework (Solo, Human-centred, AI-assisted PRactice) was introduced in prior work as a conceptual framework for structuring AI-assisted research software development \cite{Chan2026}. SHAPR explicitly recognises the interaction workspace between human researchers and generative AI systems as a central component of contemporary research practice. Within this framework, AI systems act as collaborative tools that support exploration, prototyping, and reasoning, while the human researcher maintains epistemic authority over design decisions and knowledge generation.

SHAPR should not be understood as a new research methodology. Rather, it functions as an operationalisation layer that translates the principles of Action Design Research into a workflow suitable for modern AI-assisted environments. ADR remains the underlying research method, while SHAPR provides the practical structure through which AI-assisted development activities can be organised, documented, and transformed into research knowledge.

A central premise of SHAPR is that iterative development activities can serve as a systematic source of research knowledge when properly structured and documented. In this framework, development cycles are treated not merely as engineering iterations but as knowledge-generating processes in which exploration, artefact evolution, and evaluation contribute to the accumulation of design insights.

While prior work introduced SHAPR as a conceptual framework, this paper extends the conceptual foundation introduced in prior work \cite{Chan2026} by focusing on its operationalisation, knowledge structures, and practical implementation in AI-assisted research environments. Specifically, the paper presents a structured operational workflow that connects human–AI interaction, development cycles, research artefacts, and repository-based documentation, enabling exploratory AI-assisted development to be transformed into traceable research evidence.

The paper makes four key contributions to AI-assisted research practice:

\begin{itemize}
    \item It operationalises Action Design Research for AI-assisted research environments by extending the traditional Build–Intervene–Evaluate cycle into an iterative development reasoning cycle: Explore–Build–Use–Evaluate–Learn.
    
    \item It introduces a traceable operational workflow that links human–AI interaction workspaces, development cycles, artefact evolution, and repository-based documentation, enabling AI-assisted activities to be systematically recorded as research evidence.
    
    \item It proposes SHAPR Knowledge Units (SKUs) as a mechanism for transforming development insights into reusable and progressively generalisable knowledge, supporting structured knowledge accumulation across cycles.
    
    \item It conceptualises SHAPR as both an AI-executable and system-oriented research framework, in which structured processes can be interpreted by generative AI systems and integrated with development environments, cloud storage, and version control to support scalable and traceable research workflows.
\end{itemize}

Together, these contributions position SHAPR as a framework for conducting rigorous, transparent, and scalable AI-assisted research software development while preserving human epistemic authority. Importantly, SHAPR accommodates varying levels of AI involvement, allowing researchers to balance control, learning, and automation across different research contexts.

The remainder of the paper is organised as follows. Section \ref{2sec2} introduces the core concepts of the SHAPR framework. Section \ref{2sec3} describes the relationship between research practice, artefact evolution, and knowledge generation. Section \ref{2sec4} presents the SHAPR knowledge transformation cycle. Section \ref{2sec5} introduces the operational workflow that structures AI-assisted development. Section \ref{2sec6} explains SHAPR Knowledge Units and the process of knowledge accumulation. Section \ref{2sec7} presents the integrated SHAPR operational model. Section \ref{2sec8} discusses human–AI collaboration in SHAPR practice. Section \ref{2sec9} provides practical guidance for implementation. Section \ref{2sec10} discusses implications for AI-assisted research practice. Section \ref{2sec11} concludes the paper.

\section{SHAPR Framework Overview}\label{2sec2}

The SHAPR framework (Solo, Human-centred, AI-assisted PRactice), introduced in prior work \cite{Chan2026}, provides a structured approach for organising research software development in environments where researchers increasingly collaborate with generative AI systems. Rather than introducing a new research methodology, SHAPR functions as an operationalisation layer that translates the principles of Action Design Research (ADR) into a workflow suitable for contemporary AI-assisted development environments.

In Design Science Research (DSR) and ADR, knowledge is generated through the iterative development and evaluation of artefacts \cite{Hevner2004, Sein2011}. These approaches emphasise the interaction between problem context, artefact construction, and evaluation as a mechanism for producing design knowledge. However, their practical application often assumes collaborative development settings and conventional software engineering workflows.

In contrast, many modern research software projects are conducted by individual researchers working closely with AI-assisted development tools. Generative AI systems can support code generation, debugging, design exploration, and reasoning about implementation alternatives. As a result, a substantial portion of design exploration now occurs within conversational interaction environments rather than traditional documentation processes.

SHAPR explicitly incorporates this human–AI interaction workspace as a core component of the research process. Within the framework, AI systems act as cognitive collaborators that support exploration and development, while the human researcher retains responsibility for interpretation, evaluation, and knowledge generation.

To structure this process, SHAPR distinguishes between two complementary environments:

\begin{itemize}
  \item \textbf{Interaction workspace} – where researchers explore ideas and collaborate with AI systems during design and development.
  \item \textbf{Repository workspace} – where artefacts, documentation, and research evidence are formally recorded.
\end{itemize}

A central principle of SHAPR is that interaction with AI systems becomes research evidence only when translated into artefact changes and documented within the repository. This ensures that exploratory reasoning and AI-assisted experimentation remain traceable and reproducible.

At the core of the framework is a knowledge transformation cycle consisting of the stages \textit{Explore, Build, Use, Evaluate, and Learn}. This cycle structures iterative artefact development and provides a mechanism for transforming development activities into research knowledge.

To support systematic knowledge generation, SHAPR introduces the concept of the \textit{SHAPR Knowledge Unit (SKU)}, a structured representation of an insight explaining how a design decision influences artefact behaviour. Over successive cycles, SKUs accumulate and may be synthesised into patterns and design principles, contributing to broader methodological knowledge.

Together, these elements form a coherent framework that links interaction workspaces, development cycles, artefact evolution, and knowledge documentation. Through this integration, SHAPR enables researchers to conduct rigorous, traceable, and knowledge-generating software development in AI-assisted environments.

The key concepts of the SHAPR framework are summarised below:

\begin{description}[leftmargin=5.5cm, style=nextline]
    \item[SHAPR] Framework for structuring solo, AI-assisted research software development.
    \item[Interaction Workspace] Environment for human–AI collaboration during exploration and development.
    \item[Repository Workspace] Version-controlled environment for storing artefacts and research evidence.
    \item[SHAPR Cycle] Iterative cycle: Explore → Build → Use → Evaluate → Learn.
    \item[SHAPR Knowledge Unit (SKU)] Structured insight derived from development cycles.
    \item[Knowledge Accumulation] Process through which SKUs evolve into patterns and design principles.
    \item[Research Artefact] Software system developed as part of the research process.
\end{description}

\begin{figure}[!htbp]
\centering
\includegraphics[trim=0 500 0 0, width=0.8\textwidth]{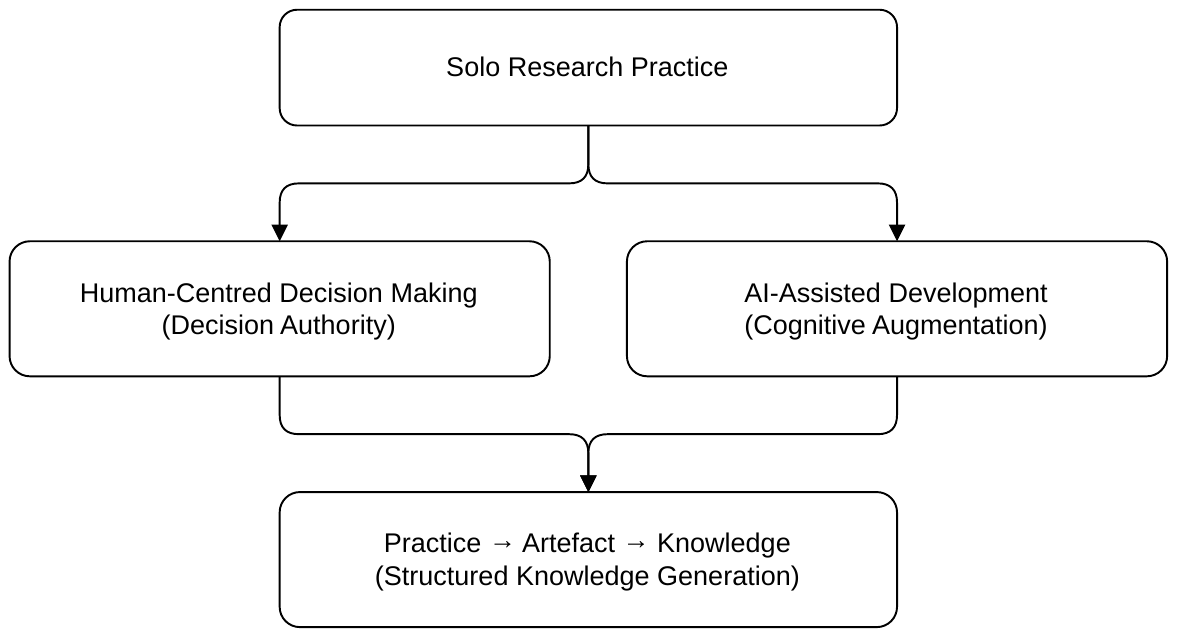}
\caption{SHAPR Conceptual Foundation. This figure illustrates the conceptual foundation of SHAPR, where solo research practice integrates human-centred decision-making and AI-assisted development. These components converge to produce practice-based artefacts that contribute to structured knowledge generation, highlighting SHAPR’s emphasis on human accountability supported by AI-assisted development.}
\label{fig:p2-fig1}
\end{figure}

\section{The Practice–Artefact–Knowledge Relationship in SHAPR}\label{2sec3}

\begin{figure}[!htbp]
\centering
\includegraphics[trim=0 580 0 0, width=0.9\textwidth]{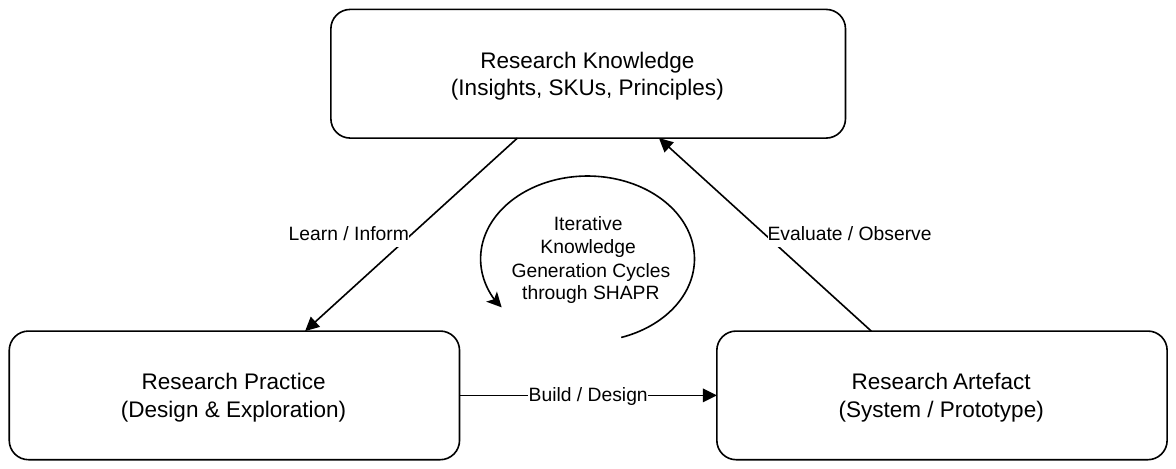}
\caption{SHAPR Iterative Cycle (ADR-Aligned). This figure presents the iterative cycle underlying SHAPR, aligned with the build–intervene–evaluate logic of Action Design Research. The cycle captures the continuous interaction between human decision-making and AI-assisted development, supported by iterative refinement, and highlights how SHAPR operationalises cyclical research practice in AI-assisted environments.}
\label{fig:p2-fig2}
\end{figure}

A central premise of the SHAPR framework is that research software development can function as a systematic process for generating research knowledge when the relationship between research practice, artefact development, and knowledge extraction is made explicit. Figure~\ref{fig:p2-fig2} illustrates this relationship as a triadic interaction between practice, artefacts, and knowledge. Rather than treating artefact construction as a purely technical activity, SHAPR conceptualises it as an epistemic process through which insights emerge from iterative development and use of research software.

This perspective aligns with the principles of design-oriented research, where artefacts embody theoretical and practical insights while simultaneously providing a medium through which new knowledge can be discovered. However, SHAPR extends this perspective by explicitly modelling how knowledge emerges from the interaction between human-centred practice and evolving artefacts in AI-assisted development contexts.

In SHAPR, the relationship between these three elements forms a knowledge-producing cycle. Research practice produces artefacts through design and experimentation; artefacts reveal insights through their behaviour, performance, and use; and these insights inform subsequent research practice. Making this relationship explicit provides a conceptual foundation for structuring AI-assisted research as a traceable and knowledge-generating process.

\subsection{Practice as Research Activity}

Within SHAPR, practice refers to the activities undertaken by the researcher during the development and exploration of research software. These activities include problem framing, design exploration, coding, testing, experimentation, interpretation of outcomes, and interaction with generative AI systems.

Practice is therefore not merely implementation work but constitutes the primary site of research activity. Through iterative development, researchers explore design alternatives, examine system behaviours, and evaluate potential solutions. Decisions made during these activities—such as selecting algorithms, structuring data flows, or refining system architectures—generate insights that extend beyond the immediate artefact.

In AI-assisted environments, generative models expand the design space by accelerating coding, prototyping, and experimentation. However, SHAPR positions these tools as assistive collaborators rather than autonomous agents. The human researcher retains responsibility for framing research questions, interpreting results, and validating knowledge claims. Research practice thus remains fundamentally human-centred, with AI augmenting exploratory capability.

\subsection{Artefacts as Knowledge-Carrying Structures}

The second element of the SHAPR relationship is the research artefact, typically represented by software systems, computational models, analytical tools, or experimental platforms developed during the research process. These artefacts embody design decisions and theoretical assumptions.

In SHAPR, artefacts are not static outcomes but evolving structures that change across iterative development cycles. As artefacts are extended, tested, and applied, they reveal behaviours and properties that may confirm, challenge, or refine the researcher’s understanding of the problem domain.

Because artefacts encode design choices, algorithms, and system architectures, they function as knowledge-carrying structures. Observing how artefacts behave under different conditions provides the empirical basis for generating research insights.

\subsection{Knowledge Extraction from Artefact Development}

The third element of the SHAPR relationship is knowledge, which represents insights derived from the interaction between research practice and artefact evolution. Unlike artefacts, this knowledge is intended to be transferable beyond a single system or implementation.

SHAPR emphasises the importance of explicitly capturing insights generated during development. To support this process, the framework introduces Structured Knowledge Units (SKUs)—documented insights that explain how specific design decisions influence artefact behaviour. Each SKU represents a traceable unit of knowledge linked to particular development activities or evaluation outcomes.

As SKUs accumulate across development cycles, recurring patterns may emerge. These patterns can be synthesised into higher-level design principles or methodological insights, contributing to the accumulation of research knowledge.

\subsection{The Practice–Artefact–Knowledge Cycle}

The interaction between practice, artefacts, and knowledge forms a continuous learning loop. Research practice generates artefacts through design and experimentation; artefacts expose insights through evaluation and use; and these insights inform subsequent development activities.

This triadic relationship transforms research software development into a structured knowledge generation process. Rather than treating artefacts as final outputs, SHAPR positions them as evolving instruments for exploration, experimentation, and learning.

The next section builds upon this conceptual relationship by introducing the SHAPR knowledge transformation cycle, which operationalises these interactions through an iterative development process aligned with Action Design Research.

\section{The SHAPR Knowledge Transformation Cycle}\label{2sec4}

To operationalise the relationship between practice, artefacts, and knowledge, the SHAPR framework introduces an iterative development process referred to as the SHAPR Knowledge Transformation Cycle. This cycle structures research software development into a sequence of activities that systematically transform exploratory practice into documented research knowledge.

Compared to traditional ADR formulations, the SHAPR cycle makes explicit the role of exploration and learning in AI-assisted environments, where iterative interaction with generative AI systems plays a central role in shaping artefact development.

\begin{figure}[!htbp]
\centering
\includegraphics[trim=0 740 0 0, width=1.0\textwidth]{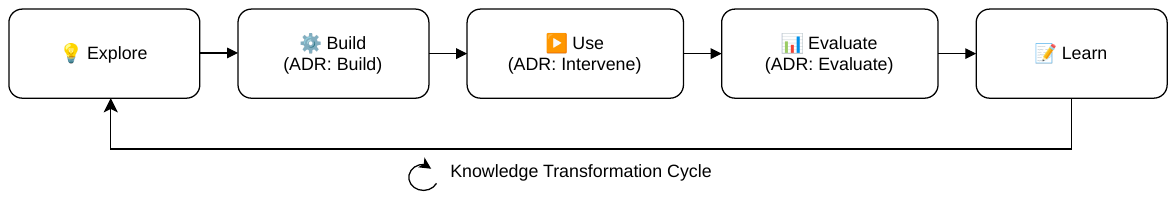}
\caption{SHAPR Conceptual Foundation. This figure illustrates the conceptual foundation of SHAPR, where solo research practice integrates human-centred decision-making and AI-assisted development. These two components converge to produce practice-based artefacts that contribute to structured knowledge generation. The model highlights SHAPR’s emphasis on maintaining human accountability while leveraging AI to support development and exploration.}
\label{fig:p2-fig3}
\end{figure}

The cycle consists of five stages: Explore → Build → Use → Evaluate → Learn, as shown in Figure~\ref{fig:p2-fig3}.

Each stage represents a distinct form of research activity, collectively enabling artefact development to function as a mechanism for generating insights.

\subsection{Explore}

The Explore stage involves investigating potential design directions, identifying limitations in the current artefact, and considering alternative approaches. Exploration may involve conceptual reasoning, literature reflection, system analysis, or interaction with generative AI systems to generate potential design options.

This stage is particularly important in AI-assisted research environments, where conversational interaction with large language models can rapidly expand the design space. However, SHAPR emphasises that exploration remains human-guided, with the researcher determining which alternatives are relevant and worth pursuing.

\subsection{Build}

In the Build stage, the selected design direction is implemented through modifications to the research artefact. This stage may involve writing code, restructuring system components, implementing algorithms, or integrating new tools.

Artefact development in SHAPR is deliberately incremental. Rather than attempting to construct complete systems in a single step, the framework encourages small, traceable modifications that can be clearly evaluated in subsequent stages.

\subsection{Use}

The Use stage corresponds to the application of the artefact within its intended context. This may involve running experiments, testing system features, analysing datasets, or simulating user workflows.

Using the artefact allows the researcher to observe how the system behaves under realistic conditions. These observations often reveal unexpected behaviours, limitations, or opportunities for improvement that were not apparent during implementation.

\subsection{Evaluate}

During the Evaluate stage, the researcher assesses the outcomes produced during the use phase. Evaluation may involve analysing system performance, examining usability, identifying errors, or comparing alternative design solutions.

This stage plays a critical role in transforming development activity into research evidence. Through systematic evaluation, the researcher determines whether the implemented design achieves the intended objectives and what new insights can be derived from the observed outcomes.

\subsection{Learn}

The final stage of the cycle is Learn, where insights derived from evaluation are explicitly captured and documented. These insights form the basis of SHAPR Knowledge Units (SKUs), representing structured observations about how particular design decisions influence artefact behaviour.

Documenting learning ensures that knowledge produced during development is not lost. Instead, insights accumulate across cycles and may later be synthesised into design patterns or methodological principles.

\section{The SHAPR Operational Workflow}\label{2sec5}

While the SHAPR knowledge transformation cycle structures the epistemic process of artefact development, researchers also require a practical workflow that integrates exploration, implementation, and evidence management. The SHAPR framework therefore introduces an operational workflow that connects the researcher’s interaction environment, development cycles, evolving artefacts, and repository-based documentation.

This workflow ensures that development activities performed during research software construction are systematically transformed into traceable research evidence. Rather than treating software development as an informal or ad hoc activity, SHAPR provides a structured mechanism through which exploratory reasoning, artefact evolution, and knowledge documentation are explicitly linked. This emphasis on traceability aligns with broader concerns in computing research regarding reproducibility and transparency \cite{Stodden2010}.

\begin{figure}[!htbp]
\centering
\includegraphics[trim=0 70 0 0, width=0.5\textwidth]{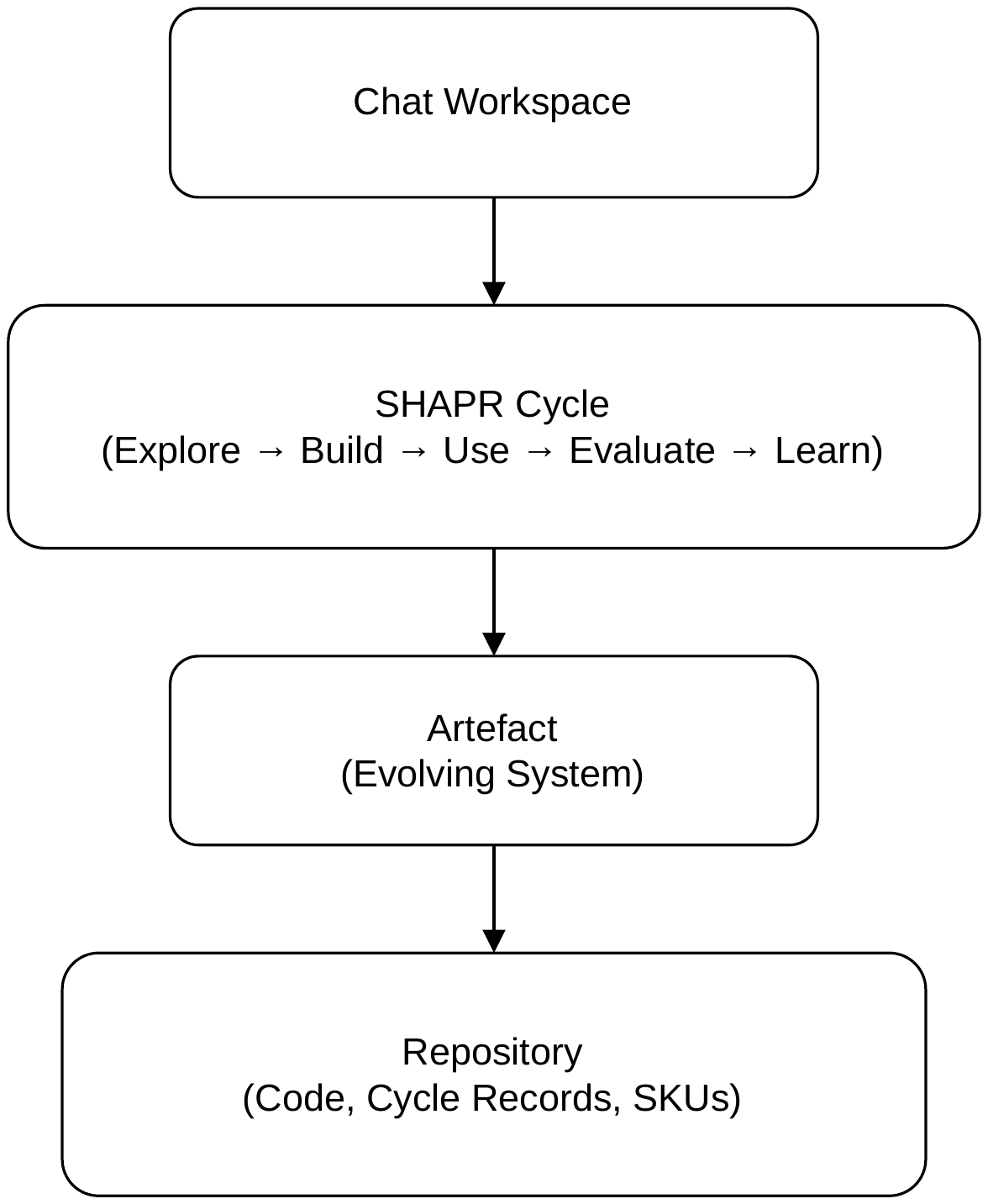}
\caption{SHAPR Workflow from Workspace to Repository. This figure illustrates the operational flow of SHAPR from the chat workspace through the SHAPR cycle to artefact development and repository storage. The SHAPR cycle (Explore–Build–Use–Evaluate–Learn) drives artefact evolution, while outputs are systematically captured in a repository containing code, cycle records, and structured knowledge units (SKUs), supporting traceability and knowledge accumulation.}
\label{fig:p2-fig4}
\end{figure}

The SHAPR operational workflow consists of four primary components: \textit{Chat Workspace → SHAPR Cycle → Artefact → Repository}. Figure~\ref{fig:p2-fig4} illustrates this end-to-end process, making explicit how human–AI interaction in the workspace leads to artefact evolution and the accumulation of structured knowledge within a repository.

Together, these components establish a traceable and reproducible workflow that supports iterative development while preserving research rigour.

\subsection{Chat Workspace: Exploration and Reasoning}

The first component of the workflow is the interaction workspace, typically represented by conversational environments such as generative AI interfaces or researcher notes. In this environment, the researcher explores ideas, evaluates design alternatives, and refines implementation strategies.

In AI-assisted research contexts, this workspace often involves interaction with large language models that support reasoning, code generation, debugging, and conceptual exploration. These interactions can significantly accelerate the exploration of design alternatives and reduce the time required to prototype new ideas.

However, SHAPR emphasises that reasoning performed in this workspace does not automatically constitute research evidence. Conversations and exploratory discussions primarily function as cognitive scaffolding. They support the researcher’s thinking process but only become part of the formal research record when translated into artefact changes and documented within the repository.

\subsection{SHAPR Cycle: Iterative Development}

The second component of the workflow is the SHAPR development cycle, which operationalises the knowledge transformation process described earlier.

Each cycle follows the sequence: \textit{Explore → Build → Use → Evaluate → Learn}.

During a cycle, the researcher investigates a design idea, implements changes to the artefact, evaluates system behaviour, and documents the resulting insights. In this way, exploratory reasoning is progressively transformed into structured development activity.

The SHAPR cycle acts as the primary unit of research progress, ensuring that development proceeds through small, traceable iterations. Each completed cycle contributes simultaneously to artefact evolution and knowledge generation.

\subsection{Artefact: Evolving Research System}

The third component of the workflow is the research artefact, typically represented by a software system under development. Examples include analytical platforms, simulation environments, experimental tools, or prototype applications.

The artefact evolves incrementally as successive SHAPR cycles introduce new features, refine system architecture, or modify algorithms. Observing artefact behaviour during testing and use provides the empirical basis for generating research insights.

In SHAPR, the artefact therefore plays a dual role:

\begin{itemize}
  \item as a technical system being developed, and
  \item as an experimental platform for generating research knowledge.
\end{itemize}

This dual role is central to SHAPR, as it links software development directly to knowledge production.

\subsection{Repository: Evidence and Knowledge Storage}

The final component of the workflow is the research repository, where development activities and insights are formally recorded. The repository typically contains source code, cycle documentation, prompts, and knowledge units derived from SHAPR cycles.

The repository serves as the primary location for research evidence, ensuring that artefact evolution and associated learning remain transparent and reproducible. A key principle of SHAPR is that development activities become part of the research record only when they are documented in the repository.

By maintaining structured records, the repository enables researchers to trace how specific design decisions influence artefact behaviour and contribute to knowledge generation.

\subsection{Traceability and Evidence}

The integration of the four workflow components enables SHAPR to maintain traceability across the entire development process. Exploratory reasoning in the workspace informs development cycles; cycles modify the artefact; artefact behaviour produces observations; and these observations are recorded and transformed into structured knowledge within the repository.

Through this workflow, research software development becomes a traceable knowledge production process rather than an opaque technical activity. The repository therefore functions not only as a storage location for artefacts but also as a historical record of the research process.

\begin{figure}[!htbp]
\centering
\includegraphics[trim=0 520 0 0, width=0.8\textwidth]{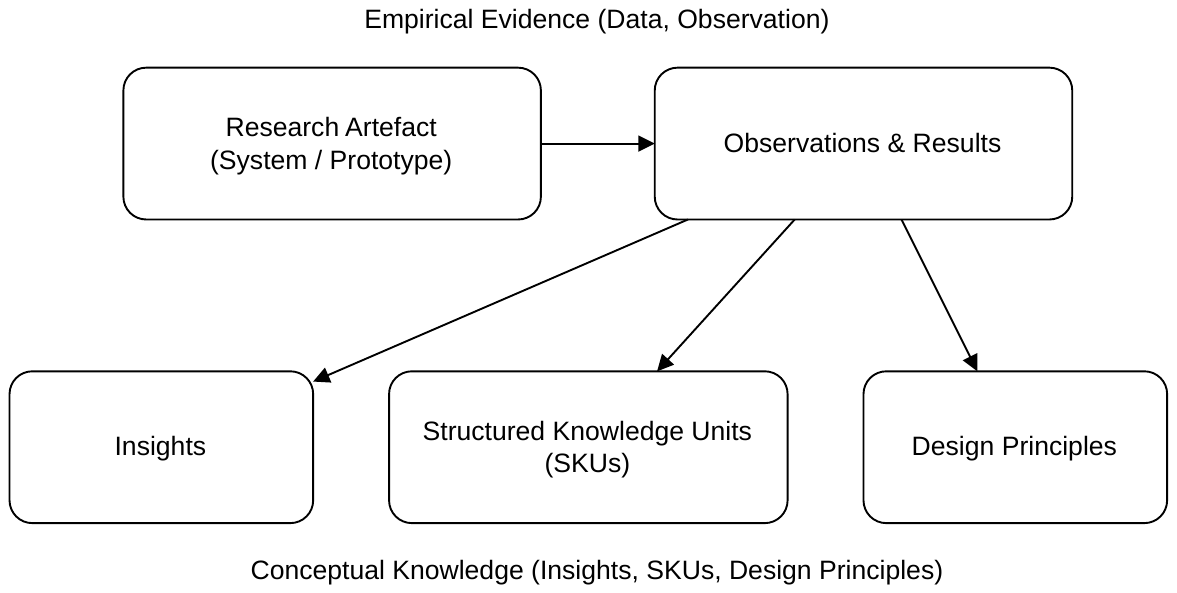}
\caption{Artefact-to-Knowledge Transformation in SHAPR. This figure illustrates how SHAPR transforms empirical evidence into conceptual knowledge. The upper layer represents empirical evidence generated through research artefacts, while the lower layer represents conceptual knowledge, including insights, structured knowledge units (SKUs), and design principles. The branching structure highlights that knowledge generation is non-linear and can occur at multiple levels of abstraction.}
\label{fig:p2-fig5}
\end{figure}

While Figure~\ref{fig:p2-fig4} illustrates how artefacts and records are generated and stored, it does not fully explain how knowledge is derived from these artefacts. This transformation is elaborated in Figure~\ref{fig:p2-fig5}.

\section{SHAPR Knowledge Units and Knowledge Accumulation}\label{2sec6}

While the SHAPR operational workflow structures the development process, a key challenge in research software development is ensuring that insights generated during artefact construction are systematically captured and transformed into research knowledge. Without deliberate documentation and reflection, valuable learning obtained during development cycles may remain implicit and be lost over time.

To address this challenge, SHAPR introduces the concept of \textit{SHAPR Knowledge Units (SKUs)} and a structured model of knowledge accumulation. This mechanism ensures that insights derived from artefact development are progressively transformed into reusable and generalisable design knowledge.

\subsection{SHAPR Knowledge Units (SKUs)}

A SHAPR Knowledge Unit (SKU) is a structured representation of an insight derived from a development cycle that explains how a specific design decision influences artefact behaviour. SKUs serve as the fundamental unit of knowledge captured within SHAPR.

Each SKU typically contains four core elements:

\begin{description}[leftmargin=3cm, style=nextline]
    \item[\textnormal{\textbf{\underline{Component}}}] \textbf{\underline{Description}}
    \item[Context] The design situation or problem addressed.
    \item[Design Decision] The specific implementation or modification applied.
    \item[Observation] The outcome or behaviour observed during evaluation.
    \item[Insight] The interpretation explaining why the outcome occurred.
\end{description}

By explicitly linking design decisions to observed outcomes, SKUs capture the causal relationship between development actions and artefact behaviour. This structured representation enables insights generated during development to be systematically recorded, analysed, and reused.

SKUs are typically extracted during the \textit{Learn} stage of the SHAPR cycle, where the researcher reflects on evaluation outcomes and identifies insights that extend beyond the immediate development task. Depending on the research context, SKUs may align with or complement established forms of knowledge representation:

\begin{description}[leftmargin=3cm, style=nextline]
    \item[\textnormal{\textbf{\underline{Method}}}] \textbf{\underline{Knowledge Unit}}
    \item[Experiment] Result
    \item[Case Study] Finding
    \item[Design Science] Design Principle
    \item[SHAPR] Structured Knowledge Unit (SKU)
\end{description}

\subsection{From Observations to Design Knowledge}

While individual SKUs represent discrete insights, knowledge accumulation in SHAPR emerges through the progressive abstraction of these insights across cycles.

SHAPR conceptualises knowledge development as a multi-level process:

\begin{description}[leftmargin=4cm, style=nextline]
    \item[\textnormal{\textbf{\underline{Stage}}}] \textbf{\underline{Knowledge Produced}}
    \item[Observation] Outcome of a development cycle.
    \item[Insight] Interpretation of the observed outcome.
    \item[SKU] Structured and contextualised insight.
    \item[Pattern] Recurring relationships across multiple SKUs.
    \item[Design Principle] Generalised knowledge applicable across systems.    
    \item[SHAPR Refinement] Adaptation or extension of the framework.
\end{description}

This hierarchy reflects the gradual transformation of practice-based observations into increasingly generalisable knowledge. As SKUs accumulate, recurring patterns may emerge, revealing consistent relationships between design decisions and system behaviour. These patterns can then be synthesised into design principles that inform future development.

\subsection{Knowledge Accumulation Across Cycles}

Knowledge accumulation in SHAPR occurs through the systematic documentation and synthesis of SKUs across successive development cycles. Each cycle contributes new observations and insights, which are recorded and stored within the repository.

As the number of cycles increases, the repository evolves into a structured knowledge base that captures both artefact evolution and associated learning. This accumulation enables researchers to identify relationships that may not be visible within individual cycles.

Importantly, knowledge accumulation may extend across multiple artefacts within a research programme. Insights derived from one artefact can inform subsequent developments, allowing knowledge to propagate across projects and contexts.

Through this process, artefact development becomes a cumulative research activity in which each cycle contributes incrementally to a broader body of knowledge.

\subsection{From Knowledge Accumulation to Framework Refinement}

At a higher level of abstraction, knowledge accumulated through SKUs and design principles can contribute to the refinement of the SHAPR framework itself. Patterns observed across artefacts and development cycles may reveal methodological insights about how AI-assisted research software development can be structured more effectively.

These insights may lead to improvements in workflow design, documentation practices, or methodological guidelines. In this way, SHAPR evolves through empirical evidence generated from practice, rather than solely through theoretical development.

The knowledge accumulation process therefore operates across multiple levels:

\begin{itemize}
  \item development cycles generate SKUs,
  \item SKUs reveal patterns,
  \item patterns produce design principles, and
  \item design principles inform framework refinement.
\end{itemize}

\subsection{Traceability and Non-Linear Knowledge Development}

Knowledge accumulation in SHAPR is not strictly linear. A single set of observations may generate multiple SKUs at different levels of abstraction, which may later be refined, combined, or generalised into broader principles. This reflects the dynamic nature of practice-based research, where knowledge evolves alongside artefacts.

The integration of evidence and traceability strengthens this process. By maintaining explicit links between actions, artefacts, and resulting knowledge, SHAPR enables researchers to trace how specific insights and SKUs were derived. This supports validation, reproducibility, and reflective learning, while facilitating the transfer of knowledge across projects and contexts.

Overall, SHAPR positions knowledge accumulation as a structured, traceable, and iterative process tightly integrated with artefact development. Through SKUs and repository-based organisation, practice-based evidence is systematically transformed into reusable and progressively refined knowledge.

\section{Integrated SHAPR Operational Model}\label{2sec7}

Building on the conceptual foundation (Section \ref{2sec3}), operational cycles (Section \ref{2sec4}), workflow and traceability (Section \ref{2sec5}), and structured knowledge accumulation (Section \ref{2sec6}), SHAPR can be understood as an integrated operational model that coordinates human decision-making, AI-assisted development, artefact evolution, and knowledge generation within a unified framework.

\begin{figure}[!htbp]
\centering
\includegraphics[trim=0 500 0 0, width=0.95\textwidth]{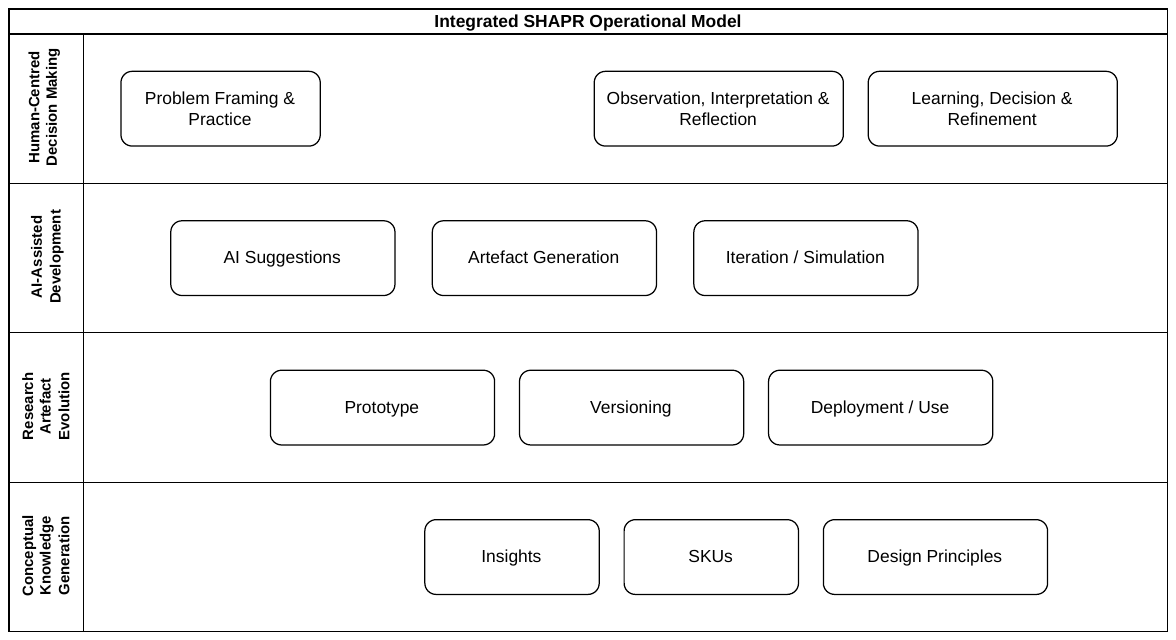}
\caption{Integrated SHAPR Operational Model. This figure presents a swimlane view of SHAPR, where the four lanes represent human-centred decision-making, AI-assisted development, research artefact evolution, and conceptual knowledge generation. Artefacts evolve through iterative cycles, generating insights, structured knowledge units (SKUs), and design principles. Evidence and traceability operate as a cross-cutting layer, linking these aspects and supporting transparency, reproducibility, and systematic refinement.}
\label{fig:p2-fig6}
\end{figure}

Figure~\ref{fig:p2-fig6} presents this integrated view as a swimlane model, where each lane represents a distinct but interconnected aspect of the SHAPR process: human-centred decision-making, AI-assisted development, research artefact evolution, and conceptual knowledge generation.

The human-centred decision-making layer captures activities such as problem framing, interpretation of results, and iterative refinement. These activities emphasise human accountability, critical judgement, and contextual understanding. While AI systems contribute to development and analysis, SHAPR maintains that responsibility for decision-making and evaluation remains with the researcher.

The AI-assisted development layer supports the execution of tasks across the SHAPR cycle, including code generation, design suggestions, debugging, and documentation. Rather than replacing human reasoning, AI acts as a collaborator that augments exploration and accelerates development while operating within the structure defined by the framework.

The research artefact evolution layer represents the incremental development of the system under study. Artefacts evolve through successive SHAPR cycles, with each iteration introducing new features, modifications, or refinements. As artefacts are developed and used, they generate empirical observations that form the basis for knowledge extraction.

The conceptual knowledge generation layer captures the transformation of empirical evidence into structured knowledge, including insights, Structured Knowledge Units (SKUs), patterns, and design principles. This layer reflects the processes described in Section \ref{2sec6}, where knowledge accumulates through iterative abstraction and synthesis.

Importantly, these four layers are not independent but operate as an interconnected system. Human decisions guide AI-assisted development; AI-supported activities contribute to artefact evolution; artefact behaviour produces observations; and these observations are transformed into structured knowledge. This knowledge, in turn, informs subsequent decisions and development cycles.

Evidence and traceability operate as a cross-cutting mechanism linking all layers. By maintaining explicit connections between decisions, artefacts, and knowledge, SHAPR enables researchers to trace how insights are derived and how artefacts evolve over time. This traceability supports reproducibility, validation, and reflective learning.

The integrated model also accommodates varying levels of AI involvement. Researchers may apply SHAPR in a predominantly human-driven manner, using AI selectively for support, or adopt more automated, agent-assisted workflows. In both cases, the framework provides a consistent structure that ensures traceability and methodological rigour.

Overall, the integrated SHAPR operational model positions research software development as a coordinated system of human judgement, AI capability, artefact evolution, and structured knowledge generation. By making these relationships explicit, SHAPR provides a foundation for conducting rigorous, transparent, and scalable research in AI-assisted environments.

\section{Human–AI Collaboration in SHAPR Practice}\label{2sec8}

Having established the integrated SHAPR operational model, it is important to clarify how human researchers and AI systems interact within this framework. The rapid advancement of generative artificial intelligence has significantly transformed the way software systems can be designed and implemented. Large language models and related AI systems are increasingly capable of generating code, proposing architectural solutions, analysing data, and assisting with technical problem solving. These capabilities raise important questions regarding the role of human researchers in software-driven research environments.

The SHAPR framework addresses this challenge by positioning artificial intelligence not as a replacement for the researcher but as a collaborative tool that augments human research practice. This perspective aligns with established principles of human–AI collaboration that emphasise human-centred design and decision-making \cite{Amershi2019, Shneiderman2022}. While AI systems can assist with exploration, development, and experimentation, the interpretation of results and the validation of research knowledge remain fundamentally human responsibilities.

In this sense, SHAPR supports an AI-augmented research practice in which generative AI expands the researcher’s capacity to explore design alternatives and accelerate development cycles, while preserving the central role of human judgement in knowledge production.

\subsection{AI as an Exploratory Partner}

A key contribution of generative AI in research software development is its ability to support rapid exploration of design ideas. Through conversational interaction, researchers can evaluate alternative architectures, generate prototype implementations, and investigate potential solutions to technical problems.

Within the SHAPR workflow, this capability primarily supports the \textit{Explore} and \textit{Build} stages of the development cycle. AI systems can assist with generating code fragments, suggesting debugging strategies, and proposing alternative design approaches. This accelerates experimentation and enables researchers to explore a broader design space than would typically be feasible.

However, SHAPR emphasises that exploration remains human-guided. The researcher determines which alternatives are relevant, evaluates the plausibility of AI-generated suggestions, and decides how these suggestions should influence artefact development.

\subsection{Human Epistemic Authority}

Although AI systems can assist with technical implementation and reasoning, they do not independently generate validated research knowledge. Within SHAPR, the human researcher retains epistemic authority, meaning that responsibility for interpreting observations, validating insights, and establishing knowledge claims remains with the researcher.

This distinction is essential for maintaining the scientific integrity of AI-assisted research. AI systems may produce plausible suggestions or explanations, but determining whether an observation constitutes meaningful knowledge requires contextual understanding, methodological reasoning, and critical evaluation.

SHAPR therefore distinguishes between artefact generation and knowledge validation. While AI contributes to artefact construction, the extraction and validation of research knowledge are grounded in human interpretation and reflection.

\subsection{Expanding the Design Space}

Generative AI systems can assist with code generation, debugging, documentation, and design exploration. When integrated within structured workflows, these systems can significantly accelerate development and expand the design space explored by researchers, as observed in recent studies of large language model-assisted programming \cite{Vaithilingam2022}.

Within the SHAPR cycle, this expanded design space enhances the exploratory capabilities of the \textit{Explore} stage and accelerates implementation in the \textit{Build} stage. By reducing the effort required to prototype new ideas, AI systems enable researchers to complete more development cycles within a given timeframe.

This increased iteration capacity strengthens SHAPR’s knowledge accumulation process, as more cycles generate a larger set of observations and insights that contribute to the formation of SKUs and design principles.

\subsection{Structured Human–AI Research Practice}

The integration of AI into research workflows introduces the risk that development activities may become opaque or difficult to reproduce, particularly when substantial portions of code or reasoning are generated through conversational interaction. SHAPR addresses this challenge by structuring the interaction between human reasoning, AI assistance, artefact development, and repository-based documentation.

By requiring that development activities and insights be captured through documented SHAPR cycles and stored within the repository, the framework ensures that AI-assisted development remains transparent, traceable, and reproducible. This structured workflow aligns AI-supported activities with methodological requirements and preserves the integrity of the research process.

\subsection{SHAPR as an AI-Guidable Framework}

An important characteristic of SHAPR is that its operational structure can be interpreted and followed by generative AI systems. Because SHAPR defines explicit cycles, documentation practices, and knowledge extraction procedures, the framework can be embedded within AI prompts, workflows, or knowledge bases.

In practice, researchers can provide SHAPR documentation—including conceptual descriptions, operational guidelines, and templates—to AI systems in order to obtain structured assistance during research software development. The AI system can then support the execution of SHAPR cycles, assist with documentation, and help identify potential knowledge units.

This capability positions SHAPR as an AI-guidable and AI-executable research framework. It enables AI systems to support not only technical implementation but also adherence to structured research processes, while preserving human-centred decision-making and epistemic authority.

\section{Practical Implementation of SHAPR}\label{2sec9}

While the preceding sections describe the conceptual and operational structure of SHAPR, its effectiveness depends on the ability of researchers to apply the framework consistently in real development environments. This section outlines practical mechanisms that enable SHAPR to be implemented in AI-assisted research software projects.

The implementation of SHAPR is grounded in three key elements:

\begin{itemize}
  \item structured development cycles,
  \item repository-based evidence management, and
  \item standardised templates for documenting insights.
\end{itemize}

Together, these elements transform exploratory software development into a traceable and reproducible research process. Detailed templates and examples are provided in Appendix A.

\subsection{Cycle-Based Development Records}

The fundamental operational unit of SHAPR is the development cycle, which captures a single iteration of exploration, implementation, evaluation, and learning. To support systematic documentation, SHAPR introduces a structured cycle record template.

Each cycle record documents the reasoning, implementation, evaluation, and learning associated with a development iteration. A typical record includes the following sections:

\begin{description}[leftmargin=3cm, style=nextline]
    \item[\textnormal{\textbf{\underline{Cycle Section}}}] \textbf{\underline{Purpose}}
    \item[Explore] Describe the design problem or opportunity investigated.
    \item[Build] Document artefact modifications and implementation details.
    \item[Use] Explain how the artefact was tested or applied.
    \item[Evaluate] Analyse outcomes, behaviour, or performance.
    \item[Learn] Summarise insights derived from the cycle.
    \item[SKU Extraction] Record reusable insights generated from the cycle.
    \item[Next Direction] Identify directions for subsequent cycles.  
\end{description}

These records ensure that development activities remain traceable and reproducible, allowing researchers to understand how artefacts evolve and how insights are generated across iterations.

A minimal template for documenting SHAPR cycles is provided in Appendix A (Section \ref{2appd}) and can be adapted to different development contexts.

\subsection{Repository-Based Evidence Management}

A second practical component of SHAPR is the use of a research repository to store artefacts, development records, and knowledge units. The repository serves as the central location for preserving research evidence and maintaining traceability.

Typical repository contents include:

\begin{itemize}
  \item source code and system components,
  \item cycle documentation,
  \item extracted SKUs,
  \item experimental results and evaluation data.
\end{itemize}

Organising these materials within a structured repository ensures that artefact evolution and associated knowledge remain transparent and reproducible.

A typical repository structure may include:

\begin{verbatim}
project_repository/
docs/
   cycles/
   insights/
src/
   system_code/
data/
   experiments/
   analysis/
design/
   prds/
   framework_notes/
\end{verbatim}

This separation of implementation artefacts and research documentation enables clear distinction between system development and knowledge generation.

\subsection{Capturing Knowledge Through SKUs}

As discussed in Section \ref{2sec6}, the extraction of SKUs is central to transforming development experience into reusable knowledge. In practice, SKUs are recorded as concise, structured entries within the repository.

Each SKU captures a reusable insight derived from a development cycle and typically includes:

\begin{description}[leftmargin=3.5cm, style=nextline]
    \item[\textnormal{\textbf{\underline{SKU Component}}}] \textbf{\underline{Description}}
    \item[Context] Situation or problem addressed.
    \item[Design Decision] Implementation approach used.
    \item[Observation] Behaviour observed during evaluation.
    \item[Insight] Explanation of why the outcome occurred.
\end{description}

This structured representation enables researchers to identify patterns across cycles and artefacts, supporting the development of design principles and methodological insights over time.

\subsection{Integrating AI Assistance into the Workflow}

Generative AI systems can support SHAPR workflows by assisting with tasks such as code generation, debugging, documentation, and design exploration. When integrated within the structured SHAPR process, AI can significantly accelerate development cycles and expand the range of design alternatives explored.

However, AI-generated outputs do not automatically constitute research knowledge. All artefact changes, observations, and insights must be evaluated by the researcher and documented through cycle records and SKU extraction. This ensures that AI-assisted development remains aligned with SHAPR’s emphasis on traceability and methodological rigour.

\subsection{Supporting AI-Guided SHAPR Workflows}

An additional advantage of SHAPR is that its structured cycles, templates, and documentation practices can be embedded within AI prompts or knowledge bases. By providing SHAPR documentation to AI systems, researchers can obtain structured assistance in applying the framework during development.

In practice, AI systems can assist with:

\begin{itemize}
  \item generating cycle records from development discussions,
  \item suggesting potential SKUs based on evaluation outcomes,
  \item identifying patterns across multiple development cycles.
\end{itemize}

To support this process, SHAPR provides reusable templates (see Appendix A), including:

\begin{itemize}
    \item cycle templates,
    \item prompt templates,
    \item reflection templates,
    \item product or project design templates.
\end{itemize}

This capability enables SHAPR to function as both a human-guided methodology and an AI-guidable framework, supporting consistent and scalable research workflows across different levels of AI involvement.

\section{Implications for AI-Assisted Research Practice}\label{2sec10}

The emergence of generative AI technologies has significantly altered the landscape of research software development. Tools capable of generating code, assisting with debugging, and proposing design alternatives can dramatically accelerate development. However, these capabilities also introduce methodological challenges related to transparency, reproducibility, and the role of human judgement in research.

The SHAPR framework addresses these challenges by providing a structured and traceable approach for integrating AI assistance into research practice while maintaining the rigour required for knowledge generation. The implications of this approach extend beyond individual tools to the design of AI-assisted research workflows, systems, and educational practices.

\subsection{Structuring AI-Assisted Research Workflows}

A key implication of SHAPR is that AI-assisted development can be organised into a reproducible and traceable research workflow. Without structured processes, development activities supported by generative AI may become difficult to interpret, verify, or replicate, particularly when substantial portions of reasoning and implementation are generated through conversational interaction.

By structuring development into explicit cycles and requiring documentation of artefacts, observations, and insights, SHAPR ensures that AI-assisted activities remain transparent and reproducible. Repository-based records further support this process by maintaining persistent links between actions, artefacts, and resulting knowledge. This enables researchers to benefit from AI capabilities while preserving methodological integrity.

\subsection{Supporting Solo Research Software Development}

Another important implication concerns the growing prevalence of solo research software development, particularly in contexts where researchers increasingly rely on AI tools to support implementation. In such environments, researchers may lack the collaborative structures and resources traditionally associated with large-scale software engineering projects.

SHAPR provides a framework that enables individual researchers to conduct systematic, traceable development while maintaining research rigour. By organising work into manageable cycles and emphasising structured documentation, the framework transforms iterative development into a disciplined process of knowledge generation. This lowers the barrier for conducting complex research software projects while maintaining academic standards.

\subsection{Enhancing Reproducibility in Software-Based Research}

Reproducibility remains a significant concern in research involving complex software systems. Traditional research outputs often focus on final artefacts or experimental results without fully documenting the processes that produced them.

SHAPR improves reproducibility by capturing development decisions, experimental outcomes, and learning across iterative cycles. Through structured documentation and repository-based organisation, the evolution of artefacts and the derivation of insights are made explicit. This traceability supports not only replication but also critical evaluation of the design and reasoning underlying research artefacts.

\subsection{SHAPR as an AI-Executable Research Framework}

A distinctive feature of SHAPR is that its operational structure can be interpreted not only by human researchers but also by generative AI systems. The framework defines explicit cycles, structured documentation practices, and knowledge extraction procedures, which can be embedded within AI prompts, workflows, or knowledge bases. As a result, SHAPR is both human-readable and AI-interpretable, enabling its use as an executable guide for AI-assisted research.

This capability supports a new form of AI-guided research methodology, in which AI systems assist not only with implementation tasks but also with adherence to structured research processes. By encoding SHAPR principles into prompts or agent-based systems, AI can guide exploration, support artefact development, structure evaluation, and facilitate knowledge extraction, while preserving human-centred decision-making.

Emerging systems such as OpenClaw illustrate how SHAPR can be operationalised as an executable research workflow. In such environments, SHAPR cycles and associated documentation structures can be embedded into AI-driven systems, enabling partial automation of research activities while maintaining traceability and methodological discipline. This positions SHAPR as a bridge between human-led research design and AI-assisted execution.

\subsection{Toward Integrated AI-Assisted Research Systems}

SHAPR is designed to be tool-independent, allowing researchers to adopt different configurations of tools depending on their objectives, expertise, and desired level of automation. AI involvement may range from lightweight, human-driven workflows to highly automated, agent-based environments. These configurations, as shown in Table~\ref{tab:ai_levels}, should be understood as illustrative examples rather than prescriptive requirements, drawing on contemporary AI-assisted development tools and environments \cite{google_gemini, openai_chatgpt, github_copilot, notebooklm}.

\begin{table}[!htbp]
\centering
\caption{Illustrative Levels of AI Involvement in SHAPR Workflows}
\label{tab:ai_levels}
\begin{tabular}{p{2cm} p{4cm} p{5.5cm}}
\hline
\textbf{Level} & \textbf{Configuration} & \textbf{Characteristics} \\
\hline

Low-AI Involvement & ChatGPT + IDE (e.g., PyCharm) + Git + Cloud Storage (e.g., OneDrive) & 
Human-driven workflow with AI used for code assistance, debugging, and conceptual support. Full control and strong learning orientation. \\

\hline

Moderate-AI Involvement & ChatGPT / Gemini + NotebookLM + GitHub + Documentation Templates & 
AI supports reasoning, summarisation, and documentation. Researcher remains central but benefits from structured AI-assisted workflows. \\

\hline

High-AI Involvement & AI-assisted development environments (e.g., GitHub Copilot, Cursor) + automated documentation tools & 
AI actively assists implementation and development cycles, accelerating iteration while maintaining human oversight and validation. \\

\hline

Agent-Based Workflows & Autonomous or semi-autonomous agents (e.g., OpenClaw-style systems) managing development cycles & 
AI systems coordinate parts of the SHAPR cycle, including implementation and documentation, with human supervision guiding direction and validation. \\

\hline

Integrated AI Research Systems & End-to-end environments combining LLMs, repositories, IDEs, and knowledge systems (e.g., NotebookLM + Gemini + cloud-based workflows) & 
Highly integrated systems supporting continuous cycles of development, documentation, and knowledge extraction, aligned with SHAPR structure. \\

\hline

\end{tabular}
\end{table}

These configurations demonstrate that SHAPR does not depend on specific tools but rather on the structure of interaction between human decision-making, AI-assisted development, artefact evolution, and knowledge documentation. Regardless of the level of automation, the framework maintains human epistemic authority while enabling flexible integration of emerging AI technologies.

Beyond individual workflows, SHAPR suggests a system-level view of research practice in which multiple components are integrated into a cohesive environment. In this view, LLM workspaces support reasoning, prompting, and documentation; integrated development environments (IDEs) support artefact construction and execution; cloud storage provides persistent access to artefacts and records; and version control systems such as Git ensure traceability of changes.

Together, these components form a cloud-connected and version-controlled research infrastructure that supports continuous iteration, structured documentation, and knowledge accumulation. Within this environment, SHAPR operates as the coordinating framework that links human decisions, AI-assisted development, artefact evolution, and conceptual knowledge generation.

This system perspective reinforces the importance of evidence and traceability as core elements of AI-assisted research. By maintaining explicit links between decisions, artefacts, and knowledge, SHAPR enables transparent and reproducible research practices, addressing long-standing challenges in computational and software-based research \cite{Stodden2010}.

\subsection{Flexible Levels of AI Involvement}

An important characteristic of SHAPR is its ability to support varying levels of AI involvement within the research process. While the framework enables AI-assisted and potentially AI-executable workflows, it does not prescribe a fixed level of automation. Instead, researchers can adapt the degree of AI involvement based on their objectives, preferences, and level of expertise.

At one end of the spectrum, SHAPR can be applied in a predominantly manual mode, where researchers retain close control over design decisions, artefact development, and knowledge extraction, using AI primarily for limited support such as code suggestions or documentation assistance. This mode is particularly valuable for learning, experimentation, and maintaining deep engagement with the research process.

At the other end, SHAPR can be implemented in more automated or agent-based environments, where AI systems assist with multiple stages of the cycle, including generation, evaluation support, and knowledge structuring. In such cases, SHAPR provides the structure that ensures these activities remain traceable and methodologically aligned.

This flexibility highlights that SHAPR is not tied to a specific level of AI capability or autonomy. Rather, it functions as a human-centred framework that accommodates a continuum of AI involvement, enabling researchers to balance control, learning, efficiency, and automation according to their needs.

\subsection{Implications for Higher Education and HDR Research}

While SHAPR was initially developed to support solo HDR research practice \cite{Chan2026}, its operationalisation in this paper reveals broader implications for higher education, research supervision, and assessment in AI-assisted environments.

\begin{table}[!htbp]
\centering
\caption{Levels of AI Involvement in SHAPR for Higher Education and HDR Contexts}
\label{tab:ai_education}
\begin{tabular}{p{2cm} p{3cm} p{3cm} p{3cm}}
\hline
\textbf{Level} & \textbf{Student Role} & \textbf{Supervisor Role} & \textbf{Assessment Focus} \\
\hline

Low-AI Involvement &
Manual development with selective AI support &
Close guidance, skill development emphasis &
Understanding, reasoning, and process transparency \\

\hline

Moderate-AI Involvement &
AI-assisted coding, documentation, and exploration &
Guided supervision, emphasis on decision-making &
Justification of design decisions and interpretation \\

\hline

High-AI Involvement &
Extensive AI-supported development workflows &
Supervisory focus on validation and critical evaluation &
Evaluation of outputs, reproducibility, and insight generation \\

\hline

Agent-Based / Automated &
AI executes parts of development cycles &
Supervisor ensures methodological rigour and integrity &
Assessment of research design, traceability, and knowledge contribution \\

\hline

Integrated AI Research Systems &
Fully integrated AI-supported research environments &
Supervisor acts as mentor for research direction and ethics &
Contribution to knowledge, originality, and methodological clarity \\

\hline

\end{tabular}
\end{table}

The concept of flexible levels of AI involvement has direct implications for higher education and HDR research. Different configurations of AI-assisted workflows correspond to different stages of learning, supervision approaches, and assessment strategies. Lower levels of AI involvement emphasise skill development, conceptual understanding, and direct engagement with implementation. In contrast, higher levels shift the focus toward critical evaluation, research design, and knowledge generation.

In HDR contexts, SHAPR provides a structured approach for managing solo research practice in AI-assisted environments. By organising development into traceable cycles and capturing insights as structured knowledge units (SKUs), the framework enables systematic documentation of research progress. This supports more transparent supervision, where supervisors can evaluate not only outcomes, such as the final artefacts, but also the reasoning, experimentation, and learning processes underlying artefact development.

The framework also has important implications for assessment. In AI-assisted environments, evaluating only final artefacts becomes insufficient, as significant portions of development may be supported by generative AI systems. SHAPR enables process-oriented assessment by capturing development cycles, decision-making, and knowledge extraction. This shifts the focus from artefact correctness to traceability, justification, and epistemic contribution.

A critical implication of AI-assisted research environments concerns the reframing of academic integrity. Traditional approaches often focus on whether students have used external assistance, including generative AI systems, in the production of their work. In AI-augmented contexts, however, such binary distinctions become increasingly difficult to define and enforce.

SHAPR supports a shift from tool-based evaluation toward process-based accountability. Rather than asking whether AI was used, the emphasis moves to whether the researcher or student can explain, justify, and trace the development process and resulting artefacts. This includes demonstrating how design decisions were made, how AI-assisted outputs were evaluated, and how insights were derived and validated.

By structuring development through documented cycles and capturing insights as structured knowledge units (SKUs), SHAPR enables transparent tracing of the research process. Academic integrity is thus grounded in epistemic responsibility, where the legitimacy of work is determined by the researcher’s ability to account for their reasoning, decisions, and knowledge contributions, regardless of the level of AI assistance involved.

More broadly, SHAPR supports a reconfiguration of the roles of students and supervisors. As AI systems increasingly assist with implementation and exploration, students are required to take greater responsibility for interpretation, validation, and knowledge construction. Supervisors, in turn, shift from direct instruction toward guiding research design, ensuring methodological rigour, and supporting critical reflection.

These implications suggest that SHAPR can serve not only as a research framework but also as a pedagogical structure for AI-assisted learning environments. Future research can explore how SHAPR-informed workflows support student learning, supervision practices, and assessment design across different levels of education.

\subsection{Toward Practice-Centred Research Methodologies}

More broadly, SHAPR supports a reconfiguration of the roles of students and supervisors. As AI systems increasingly assist with implementation and exploration, students take greater responsibility for interpretation, validation, and knowledge construction. Supervisors, in turn, shift from direct instruction toward guiding research design, ensuring methodological rigour, and supporting critical reflection.

These implications suggest that SHAPR can function not only as a research framework but also as a pedagogical structure for AI-assisted learning environments. Future research can explore how SHAPR-informed workflows support student learning, supervision practices, and assessment design across different levels of education.

\section{Conclusion}\label{2sec11}

The growing capabilities of generative artificial intelligence are rapidly transforming the practice of research software development. While these technologies enable faster experimentation and implementation, they also introduce challenges related to methodological structure, traceability, and the preservation of human judgement in knowledge generation.

This paper presented SHAPR as a structured framework for operationalising Action Design Research in the context of solo, AI-assisted research software development. The framework conceptualises research software development as a knowledge-generating process by explicitly linking research practice, artefact evolution, and knowledge extraction.

The paper makes four key contributions. First, it operationalises ADR for AI-assisted environments through an iterative development reasoning cycle consisting of Explore, Build, Use, Evaluate, and Learn. Second, it introduces a traceable operational workflow that connects human–AI interaction, development cycles, artefact evolution, and repository-based documentation. Third, it proposes SHAPR Knowledge Units (SKUs) as a mechanism for transforming development insights into reusable and progressively generalisable knowledge. Fourth, it positions SHAPR as an AI-executable and system-oriented framework that integrates human-centred decision-making, AI-assisted development, and supporting infrastructure to enable scalable and reproducible research workflows.

A key feature of SHAPR is its emphasis on human epistemic authority within AI-assisted environments. While generative AI systems can significantly expand the design space and accelerate development, the interpretation, validation, and generalisation of knowledge remain the responsibility of the human researcher. In this way, SHAPR supports AI-augmented research practice while preserving methodological rigour and accountability.

Importantly, SHAPR is designed to be tool-independent in an environment where AI technologies are evolving rapidly. New tools, platforms, and agent-based systems continue to emerge, offering different capabilities for reasoning, development, and knowledge management. SHAPR accommodates this dynamic landscape by enabling researchers to adopt different configurations of tools and varying levels of AI involvement, ranging from human-driven workflows to highly automated, AI-executable systems. This flexibility allows the framework to support diverse research and educational objectives while remaining robust to changes in the underlying technology ecosystem.

By structuring development cycles, enabling systematic knowledge capture, and supporting traceability across the research process, SHAPR provides a practical foundation for transforming iterative software development into a rigorous and reproducible research activity. The framework is particularly relevant for researchers who increasingly rely on AI tools to support complex, software-driven investigations.

More broadly, SHAPR contributes to the development of AI-assisted research methodologies that are structured, transparent, and scalable. By accommodating varying levels of AI involvement and enabling integration with development environments, repositories, and AI systems, SHAPR provides a flexible foundation for future research practice.

Future work will extend SHAPR across multiple research software artefacts, domains, and educational contexts to evaluate its effectiveness and generalisability. In particular, further research can explore its application in HDR supervision, AI-assisted learning environments, and assessment design, as well as its integration with emerging AI tools and agent-based systems. As AI technologies continue to evolve, SHAPR provides a stable conceptual and operational foundation for investigating how human-centred and AI-assisted research practices can co-evolve, enabling a broad and evolving research agenda.

\section*{Appendix A. Applying SHAPR in LLM-Supported Research Workflows}\label{2appd}

This appendix provides a practical guide for applying SHAPR in LLM-supported research environments. It demonstrates how researchers can use SHAPR as an AI-readable and executable framework by embedding its structures, cycles, and templates into interactions with large language models (LLMs).

\subsection*{A.1 SHAPR as an AI-Readable Framework}

SHAPR is designed to be both human-readable and AI-interpretable. Its structured cycles (Explore–Build–Use–Evaluate–Learn), explicit artefacts, and documentation practices enable it to be directly incorporated into LLM workflows. By providing SHAPR documentation to an AI system, researchers can guide the AI to follow consistent research processes aligned with SHAPR principles.

This enables a workflow in which researchers:
\begin{itemize}
    \item Provide SHAPR framework documents to an AI system,
    \item Instruct the AI to follow SHAPR cycles,
    \item Develop artefacts collaboratively with AI assistance,
    \item Document cycles and extract structured knowledge,
    \item Accumulate reusable knowledge for future research.
\end{itemize}

\subsection*{A.2 Typical SHAPR–LLM Workflow}

A typical workflow using SHAPR in an LLM workspace is as follows:

\begin{enumerate}
    \item Upload SHAPR framework documents (e.g., Papers 1–3, templates, and notes).
    \item Initialise the AI with instructions to follow SHAPR principles.
    \item Begin the Explore phase by defining the problem and research objectives.
    \item Progress through Build, Use, Evaluate, and Learn stages with AI assistance.
    \item Document each cycle, including artefacts, observations, and reflections.
    \item Extract Structured Knowledge Units (SKUs) and update the repository.
\end{enumerate}

\subsection*{A.3 Example Initialisation Prompt}

The following prompt can be used to initialise an LLM for SHAPR-based research:

\begin{verbatim}
You are assisting in a research project using the SHAPR framework 
(Solo Human-Centred and AI-Assisted Practice).

Follow these principles:
- Maintain human-centred decision-making
- Support AI-assisted development
- Structure work using SHAPR cycles:
  Explore → Build → Use → Evaluate → Learn
- Help document each cycle clearly
- Assist in extracting Structured Knowledge Units (SKUs)
- Ensure traceability between decisions, artefacts, and knowledge

Your role is to support the researcher while preserving their 
decision authority and ensuring alignment with SHAPR.
\end{verbatim}

\subsection*{A.4 Cycle Execution Prompt Template}

For each SHAPR cycle, the following prompt structure can be used:

\begin{verbatim}
We are in the [STAGE] phase of the SHAPR cycle.

Context:
[Describe current problem, artefact, or goal]

Tasks:
- Guide the next steps for this stage
- Suggest artefact development or refinement
- Identify what should be observed or evaluated
- Help structure outputs and documentation

Output:
- Clear next steps
- Suggested artefact changes
- Key observations to capture
- Potential SKUs to extract
\end{verbatim}

\subsection*{A.5 SKU Extraction Prompt Template}

\begin{verbatim}
Based on the following observations and results:

[Insert observations]

Please:
- Identify key insights
- Convert them into Structured Knowledge Units (SKUs)
- Indicate possible reuse contexts
- Suggest whether any design principles can be derived
\end{verbatim}

\subsection*{A.6 Repository Structure (Example)}

A SHAPR repository may be structured as follows:

\begin{verbatim}
project_root/
├── artefacts/
│   ├── prototype_v1/
│   ├── prototype_v2/
│
├── cycles/
│   ├── cycle_01.md
│   ├── cycle_02.md
│
├── knowledge/
│   ├── skus.json
│   ├── design_principles.md
│
├── prompts/
│   ├── initialisation.txt
│   ├── cycle_template.txt
│
└── README.md
\end{verbatim}

\subsection*{A.7 Toward AI-Executable Research Systems}

The structured nature of SHAPR enables its integration into AI-driven environments and research tools. Systems such as OpenClaw illustrate how SHAPR cycles, prompts, and documentation practices can be embedded into AI-assisted workflows, enabling partial automation of research activities while maintaining traceability and methodological rigour.

This suggests a future in which SHAPR functions not only as a conceptual framework but also as a practical foundation for AI-executable research systems, supporting scalable, reproducible, and knowledge-generating research practices.

\subsection*{A.8 Integrated SHAPR Development Environment}

SHAPR workflows can be further simplified and strengthened by integrating LLM workspaces with development environments, cloud storage, and version control systems. In such a setup, the LLM (e.g., ChatGPT) is used for guidance, prompting, and documentation, while an integrated development environment (IDE) such as Visual Studio Code or PyCharm is used for artefact development and execution.

Cloud storage (e.g., OneDrive, Google Drive, or cloud-based repositories) enables persistent storage of artefacts, cycle records, and knowledge outputs, ensuring accessibility and continuity across sessions. Version control systems such as Git provide systematic tracking of changes to code, documents, and knowledge artefacts, supporting traceability and reproducibility.

In this integrated environment, SHAPR cycles can be executed seamlessly:
\begin{itemize}
    \item The LLM guides the researcher through SHAPR stages and documentation.
    \item The IDE supports artefact development, testing, and refinement.
    \item Cloud storage maintains structured records of cycles, artefacts, and knowledge.
    \item Git captures version histories, enabling traceability of decisions and changes.
\end{itemize}

This integration reduces friction in applying SHAPR and reinforces its emphasis on structured, traceable, and iterative research practice. It also supports the development of scalable and collaborative research workflows, where artefacts and knowledge can be shared, reused, and extended across projects and teams.

In this sense, SHAPR can be viewed as a cloud-connected and version-controlled research workflow that integrates human judgement, AI assistance, and persistent knowledge accumulation.

\end{document}